\documentclass[conference]{IEEEtran}
\IEEEoverridecommandlockouts
\usepackage{cite}
\usepackage{amsmath,amssymb,amsfonts}
\usepackage{booktabs}
\usepackage{graphicx}
\usepackage{textcomp}
\usepackage{xcolor}
\usepackage{orcidlink}
\usepackage{tikz}
\usepackage{hyperref}
\hypersetup{hidelinks} 
\usetikzlibrary{positioning,calc}

\newcommand{\agent}[1]{\textsc{#1}} 
\newcommand{\Section}[1]{Section~\ref{#1}}
\newcommand{\Figure}[1]{Fig.~\ref{#1}}
\newcommand{\Table}[1]{Table~\ref{#1}}

\begin{document}

\title{Admitted Is Not Effective: Lessons from a Notary-Gated Retrospective Loop in an Agentic Pipeline}

\author{\IEEEauthorblockN{Lo {Gullstrand Heander}\,\orcidlink{0000-0002-0695-4580}}
\IEEEauthorblockA{\textit{Lund University}\\
Lund, Sweden \\
lo.gullstrand\_heander@cs.lth.se}
\and
\IEEEauthorblockN{Emma Söderberg\,\orcidlink{0000-0001-7966-4560}}
\IEEEauthorblockA{\textit{Lund University}\\
Lund, Sweden \\
emma.soderberg@cs.lth.se}
\and
\IEEEauthorblockN{Christofer Rydenfält\,\orcidlink{0000-0003-1495-8263}}
\IEEEauthorblockA{\textit{Lund University}\\
Lund, Sweden \\
christofer.rydenfalt@design.lth.se}
}

\maketitle

\begin{abstract}
Agentic software development pipelines with frozen prompts can repeat the same mistakes across pull requests (PRs), yet ungoverned additions to instruction files grow without bound and do not generally improve task success. We present a retrospective loop for a four-stage large language model (LLM) pipeline in which a \agent{Retrospective} agent proposes edits to learned rules every ten PRs and a \agent{Notary} admits, merges, or rejects them while recording each decision in a provenance ledger. In a pre-registered study of 880 PRs on two Python repositories, the loop ran end to end and the \agent{Notary} admitted 55 of 89 proposals, rejecting 8 that cited nonexistent evidence. 

However, held-out review comments and hidden-test resolution did not improve after ten retrospectives. Tracing each step of the loop, we found that sparse review signal (0.45 comments/PR) left little to learn from and that rules admitted on their stated evidence did not show measurable held-out effects. We derive six lessons and argue that governed loops should admit and retire every learned rule as a testable hypothesis.
\end{abstract}

\begin{IEEEkeywords}
LLM agents, agentic software engineering, self-improving agents, retrospectives, code review
\end{IEEEkeywords}

\section{Introduction}
\label{sec:intro}

Agentic development pipelines with frozen prompts repeat the same mistakes in pull request (PR) after PR. The common remedy is manual: a developer notices a recurring mistake, stops the pipeline, adds a lesson to an instruction file such as \texttt{AGENTS.md} or \texttt{CLAUDE.md}, and runs it again. Whoever writes them, these files keep growing. A study of their commit histories found that they gain 4.9 net instructions per commit while older instructions are rarely deleted~\cite{chakrabarti2026claudemd}, and an evaluation on coding tasks found that they raise inference cost by over 20\% on average without generally improving task success~\cite{gloaguen2026agentsmd}. Deciding which lessons to keep is hard for human teams as well, whose retrospective discussions can reflect strong opinions rather than evidence~\cite{lehtinen2017retrospectives}. Automating the remedy, so that agents edit their own instructions, adds further risks, such as gaming the reviewer, encoding chance failures, and accumulating low-impact rules. These risks motivate an independent admission gate.

An admission gate can admit, merge, and retire working rules while keeping them auditable by humans~\cite{voria2026instructions}. However, a gate in a live project has no score that says whether an edit helped. Hidden tests rarely exist outside benchmarks, so the gate must judge from noisy proxies such as review comments, token spend, guard stops, and visible-test outcomes.

In this paper, we present a \agent{Notary}-governed retrospective loop (\Section{sec:loop}) for a four-stage large language model (LLM) agentic pipeline: \agent{Planning}, \agent{Implementation}, \agent{Review}, and \agent{Fix}. Every $N=10$ PRs, a \agent{Retrospective} agent proposes bounded edits (\texttt{add}, \texttt{replace}, \texttt{delete}, and \texttt{merge}) to the learned rules of the \agent{Planning}, \agent{Implementation}, and \agent{Fix} stages, and a \agent{Retrospective Notary} gates them while recording every decision in an append-only provenance ledger.

Our pre-registered experiment across 880 PR runs on \texttt{sqlglot} and \texttt{dvc} (\Section{sec:first-run}) shows that the governed loop runs end to end and the \agent{Notary} filters unsupported rule edits (accepting 55 of 89 and rejecting 8 that cited nonexistent comments). Held-out outcomes (review comments per PR and hidden-test resolution) did not improve measurably after ten retrospectives (\Table{tab:results}). Verbatim traces of every decision and LLM call show that the evidence a rule cited at admission did not translate into a measurable held-out effect. We therefore argue that a governed loop should treat every learned rule as a testable hypothesis, admitted and retired on measured effects. We contribute:
\begin{itemize}
\item A \agent{Notary}-governed retrospective loop and provenance ledger that separate observation, proposal, and admission in a multi-stage agentic pipeline (\Section{sec:loop}).
\item A pre-registered 880-PR experiment showing that the \agent{Notary} admitted rules on their stated evidence while held-out outcomes stayed flat, with a 480-PR uptake check and thematic coding of the 48 learned rules tracing where the signal thinned (\Section{sec:first-run}, \Figure{fig:chain}).
\item Six lessons, led by review-signal starvation ($0.45$ comments/PR), and a vision of learned rules as testable hypotheses with the controls needed to evaluate it (\Section{sec:pitfalls}, \Section{sec:directions}).
\end{itemize}

\section{Related Work}
\label{sec:related}

Methods that let LLM agents learn from experience keep lessons as text, storing reflections on failed trials~\cite{shinn2023reflexion}, extracting cross-task insights~\cite{zhao2024expel}, or distilling trajectories into playbooks~\cite{zhang2026ace}. SkillOpt is closest in edit mechanism, applying bounded add, delete, and replace edits to a skill file only when held-out validation improves~\cite{yang2026skillopt}. Variants for software development pipelines similarly optimize role prompts~\cite{shen2025textualfeedback} or rule pools~\cite{jiang2026ruleevolve}, and optimizer-side memory records each edit's rationale and observed outcome~\cite{yu2026sim}. These methods judge each edit by a score, such as task success or a validation result, which an agentic software development pipeline in a live project does not have.

Coding-agent studies report small, mixed, or null effects when the score that shows whether an edit helped is weak or held out. Treude and Baltes let Claude Code propose post-session harness edits; across seven informative evaluations, two eliminated the targeted action, one reduced it, and four showed no reduction~\cite{treude2026harnesses}. Cortex gated rules by repeated runs and found nothing to learn from a real open-source project~\cite{giampieri2026cortex}. Tomar showed that standard ``textbook'' changes recover about 90\% of a self-improving loop's gain and beat released evolved agents on held-out SWE-bench Verified~\cite{tomar2026textbook}. Consistent with these studies, our first run found null held-out gains. We support Treude and Baltes's call for validation and retirement of learned rules, adding a multi-stage setting with a frozen \agent{Review} agent and a provenance ledger.

To mitigate the lack of a clear score, recent methods probe or contrast executions. Shepard and Albrecht patch repository guidance using synthetic bug-fix probes~\cite{shepard2026proberefine}. Harness$^2$ runs reference and candidate harnesses on the same task under matched configurations, retaining lessons supported by behavioral contrasts that guide composition rather than verify correctness~\cite{xu2026harness2}. Our Run-1 \agent{Notary} checks rules statically, deferring re-execution to held-out tasks. \Section{sec:pitfalls} adapts windowed re-execution so admission is not blind to target metrics.

Accumulated instructions can also hurt. Self-picked lessons match no playbook (53.8 vs.\ 54.1)~\cite{pham2026playbooks}, propagate chance failures~\cite{nakayashiki2026lessonsluck}, fail when applied out of context~\cite{chen2026memdevbench}, and outlive their target defects~\cite{hassanzadeh2026lessonsoutlive}. Instruction files are meant to restore human governance over agents~\cite{voria2026instructions}, an aim that a loop which only adds rules works against. These risks motivate our budget, deletion, and evidence checks (\Section{sec:loop}). \Section{sec:first-run} shows that they alone deleted no rules.

\section{The Governed Retrospective Loop}
\label{sec:loop}

\begin{figure}[t]
\centering
\begin{tikzpicture}[
  font=\scriptsize,
  stage/.style={draw, rounded corners=2pt, minimum height=6mm,
                minimum width=12mm, align=center},
  frozen/.style={stage, fill=black!12, dashed},
  flow/.style={-latex},
  node distance=3.5mm]
  \node[stage] (plan) {\agent{Planning}};
  \node[stage, right=of plan] (impl) {\agent{Impl.}};
  \node[frozen, right=of impl] (rev) {\agent{Review}\\(frozen)};
  \node[stage, right=of rev] (fix) {\agent{Fix}};
  \draw[flow] ([xshift=-5mm]plan.west) node[left] {issue} -- (plan);
  \draw[flow] (plan) -- (impl);
  \draw[flow] (impl) -- (rev);
  \draw[flow] (rev) -- (fix);
  \draw[flow] (fix.east) -- ++(6mm,0) node[right] {PR};
  \coordinate (bus) at ($(plan.south |- impl.south)+(0,-3mm)$);
  \draw (bus) -- (bus -| fix.south);
  \foreach \n in {plan, impl, fix} \draw[flow] (bus -| \n.south) -- (\n.south);
  \node[stage, below=8mm of fix] (ra) {\agent{Retro}};
  \node[stage, below=8mm of impl] (no) {\agent{Notary}};
  \node[stage, below=8mm of plan] (led) {Ledger};
  \draw[flow, dashed] ([xshift=3mm]fix.east) -- ([xshift=3mm]fix.east |- ra.east)
    node[pos=0.55, right, align=left] {evidence,\\every $N$\\PRs} -- (ra.east);
  \draw[flow] (ra) -- node[above] {proposals} (no);
  \draw[flow] (no) -- (led);
  \draw[flow] (no.north) -- node[right] {rules $v_{k+1}$} (no.north |- bus);
\end{tikzpicture}
\caption{The governed retrospective loop. Every $N$ PRs, \agent{Retrospective} proposes rule edits from the review evidence; \agent{Notary} judges them, writes
the next learned rules of \agent{Planning}, \agent{Implementation}, and \agent{Fix}, and records each
decision in the ledger. \textbf{Takeaway:} the loop can change how the
pipeline plans, implements, and fixes, but never the reviewer that produces
its evidence.}
\label{fig:loop}
\end{figure}
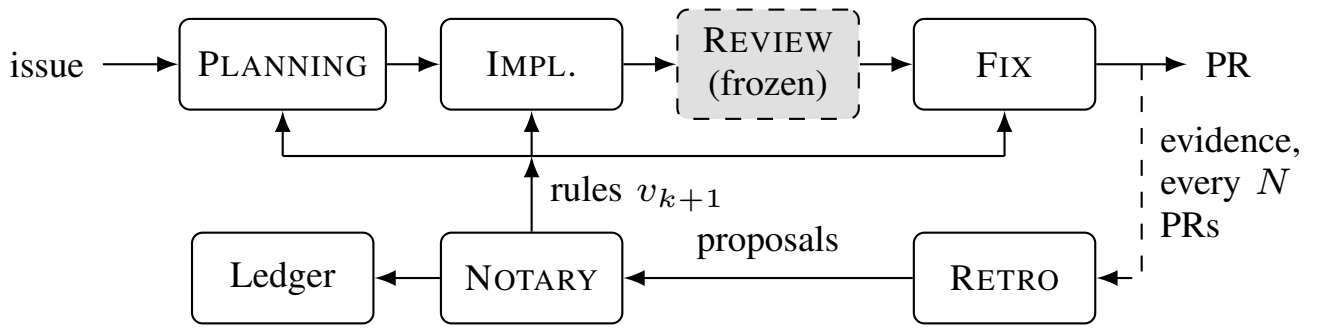

The loop separates three roles: \agent{Review} observes, \agent{Retrospective} proposes, and \agent{Notary} decides (\Figure{fig:loop}). All agents run on one configured LLM (\texttt{Qwen3.6-35B-A3B} in Run~1, \Section{sec:first-run}) in a custom harness that logs every LLM and tool call verbatim. Only the \agent{Planning}, \agent{Implementation}, and \agent{Fix} instructions change. The \agent{Review} prompt stays frozen so its comments remain comparable across windows, while the \agent{Notary} is the governance mechanism this paper argues for.

\subsection{Pipeline and Learned Rules}

The pipeline resolves one issue per PR in four tool-using LLM agent stages. \agent{Planning} writes a plan, \agent{Implementation} edits code, \agent{Review} comments on the diff, and \agent{Fix} addresses or declines each comment. \agent{Review}'s prompt is frozen and hidden from every other agent. Agents see only the issue description and the repository; an evaluation step outside the pipeline scores each PR with \emph{hidden tests} that no agent sees (\Section{sec:first-run}). Each editable prompt splits into a fixed \emph{contract} (i.e., role, inputs, schema, and study constraints) and binding \emph{learned rules} holding working policy with stable identifiers and provenance links. Learned rules are versioned, with $v_k$ the version after $k$ retrospectives; $v_0$ holds 36 author-written seed rules (1,070 tokens across the three stages, 227--440 per stage; seed limit 1,500 per stage), which the loop may replace, merge, or delete without special status.

To illustrate what learned rules contain, we coded the 48 non-seed rules at $v_{10}$ across the four trajectories of Run~1 (\Section{sec:first-run}). All three authors inductively built a codebook from \texttt{sqlglot-a}. The first author applied it to all four trajectories, the second author reviewed every assignment, and the two resolved disagreements in a joint session that merged 20 tags into 10. Each rule carries one \emph{context} code: generic (23), project-specific (23), or Python-specific (2), although the split differs by repository: generic rules dominate in \texttt{dvc} (19 of 22) and project-specific rules in \texttt{sqlglot} (22 of 26). Each rule also carries one or more \emph{topic} codes: {implementation consistency} (17), {pre-test validation} (12), {interface or architectural consistency} (10), {error handling} (6), {API and signature grounding} (6), {implementation conventions} (6), and {dead-code removal} (3). For instance, \texttt{dvc-b} \texttt{IMPL-016} (project-specific, implementation conventions) prefers ``tree abstraction methods (e.g., \texttt{tree.walk()}, \texttt{tree.open()}) over direct \texttt{os} module calls.'' Similarly, \texttt{sqlglot-a} \texttt{PLAN-011} (project-specific, interface or architectural consistency) adds new dialect keywords ``only to that dialect's components rather than the base components.''

\subsection{Retrospective Agent and Notary}

After every $N=10$ PRs, \agent{Retrospective} inspects issue titles, plans, diff statistics, review comments, \agent{Fix} outcomes, agent-run tests, current rules, prior edits, rejected proposals, and stage contracts and tools, without seeing hidden tests, the \agent{Review} prompt, or study metrics. It proposes \texttt{add}, \texttt{replace}, \texttt{delete}, or \texttt{merge} edits, citing at least two review comments, or one comment and one ledger entry.

The \agent{Notary} gates $v_k \to v_{k+1}$ and has three parts: deterministic checks, an LLM judge, and a compaction step. The deterministic checks verify that cited identifiers exist, reject proposals sharing any six-word sequence with the frozen \agent{Review} prompt (so rules cannot copy the \agent{Review} agent's criteria) and cap edits at three per stage and window, analogous to SkillOpt's learning rate~\cite{yang2026skillopt}. The LLM judge then rejects contract contradictions, rejects or converts rule contradictions into \texttt{replace}, converts duplicates into \texttt{replace} or \texttt{merge}, checks that cited comments demonstrate the problem across two changes (in a dry run, it rejected a planted unsupported proposal in 5/5 repeats), and checks specificity. Finally, compaction enforces a 3,000-token per-stage budget by merging or deleting rules.

\subsection{Provenance Ledger and Intended Safeguards}

An append-only ledger logs every proposal's window, operation, rule ID, stage, before/after text, cited evidence, decision, and reason. The next window's \agent{Retrospective} sees the previous window's rejected proposals with the \agent{Notary}'s reasons, so that it does not repeat them, similar to SkillOpt's buffer of failed edits~\cite{yang2026skillopt}. The design's safeguards target five risks from \Section{sec:related}. The frozen, hidden \agent{Review} prompt, leakage check, and hidden tests limit \emph{sensor gaming}. The 3,000-token ceiling and \texttt{delete} and \texttt{merge} operations bound \emph{growth}~\cite{chakrabarti2026claudemd}. The two-comment rule and evidence check filter \emph{noise}~\cite{nakayashiki2026lessonsluck}, and uptake probes test \emph{ignored rules}. The design leaves the fifth risk, \emph{self-judgment}, unmitigated, since one LLM both proposes and judges~\cite{pham2026playbooks}. LL~3 proposes admitting rules on measured outcomes instead. Sections~\ref{sec:first-run} and~\ref{sec:pitfalls} assess which safeguards held.

\section{First Run: The Loop Works, the Pipeline Does Not Improve}
\label{sec:first-run}

We evaluated the retrospective loop in a pre-registered run (\href{https://doi.org/10.5281/zenodo.23978257}{DOI:10.5281/zenodo.23978257}) on four arm64 DGX Sparks serving a local open-weight LLM for reproducibility~\cite{baltes2026llmguidelines}, \texttt{nvidia/Qwen3.6-35B-A3B-NVFP4}. Chronological \texttt{SWE-rebench} tasks ran in validated arm64 containers on \texttt{tobymao/sqlglot} and \texttt{iterative/dvc} across two seeded trajectories (\texttt{a}, \texttt{b}) per repository: 100 loop issues per trajectory over 10 windows of $N=10$ ($v_0 \to v_{10}$, 400 PRs, 40 retrospectives), then 60 held-out issues per repository run under both $v_0$ and $v_{10}$ per trajectory (paired, 480 PRs; 880 total, 0 infrastructure drops; 1 unparsed review in \texttt{dvc-b} $v_0$ excluded from comment counts). Each \texttt{SWE-rebench} task pairs a GitHub issue description and the repository at the issue's base commit with the tests from the PR that fixed it~\cite{badertdinov2025swerebench}. These hidden tests ran only at evaluation, in a fresh container; a PR resolved its issue when all tests that the original fix made pass (\texttt{FAIL\_TO\_PASS}) passed and all previously passing tests (\texttt{PASS\_TO\_PASS}) still passed.

\begin{figure}[t]
\centering
\includegraphics[width=\columnwidth]{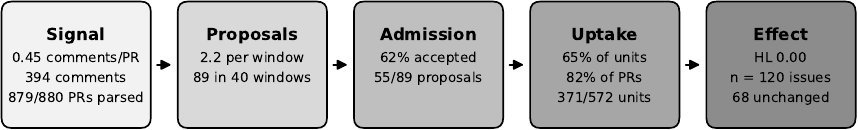}
\caption{Signal-to-effect chain in Run~1, from review comments per parsed PR to paired held-out Hodges-Lehmann (HL) comment shift. Uptake is measured on a planted probe rule, not on the admitted rules. \textbf{Takeaway:} governance and uptake of a file-level probe rule worked as designed, yet produced no measurable held-out improvement, which we attribute to sparse review signal and outcome-blind admission (\Section{sec:pitfalls}).}
\label{fig:chain}
\end{figure}

\begin{table*}[t]
\centering
\caption{Summary of Run-1 governance mechanics, rule evolution, uptake, and
held-out outcomes across four trajectories (880 PRs). \textbf{Takeaway:} the
governed loop ran end to end and agents followed a file-level probe rule in most files, yet
held-out review, test, and cost metrics showed no downstream improvement from
$v_0$ to $v_{10}$.}
\label{tab:results}
\footnotesize
\begin{tabular}{@{}lp{0.58\textwidth}p{0.24\textwidth}@{}}
\toprule
Measure & Run-1 Result & Interpretation \\
\midrule
Loop mechanics &
880 PRs completed (0 dropped); 40 retrospectives ($v_0 \to v_{10}$); 1 unparsed review (\texttt{dvc-b} $v_0$). &
Governed loop runs end to end. \\
\agent{Notary} gate &
55/89 accepted (62\%: 46 add, 5 merge, 4 replace, 0 delete); 34 rejected (8 nonexistent comment IDs in pre-check, 25 unsupported/single-instance evidence, 1 contradiction). &
Provenance and evidence gates hold. \\
Rule growth &
Combined 3-stage tokens 1,070 $\to$ 1,528--1,801 (peak section 820 of 3,000 budget, \texttt{sqlglot-b} Impl.); 0 deletions; 136/144 seed instances kept unmerged and unreplaced. &
Budget never binds; no retirement. \\
Rule content ($v_{10}$) &
48 non-seed rules (11, 15, 12, 10 across \texttt{sqlglot-a/b}, \texttt{dvc-a/b}): 23 generic, 23 project-specific, 2 Python; 31/48 (65\%) say ``verify'' (\Section{sec:loop}). &
Generic or domain-level practice. \\
Uptake check &
File-level probe (\texttt{\# Changed:} once per modified file) followed in 371/572 files (65\%; start 65\%, end 65\%) and 197/240 PRs (82\%) at least once, 101/240 (42\%) in every file; function-level probe (\texttt{Changed:} docstring per added or modified function) in 130/789 functions (16\%; start 22\%, end 12\%) and 85/240 PRs (35\%) at least once, 4/240 (2\%) in every function; baselines 0/577, 0/786. &
A once-per-file rule is followed far more often than a per-function rule; end placement lowered only the latter. \\
Held-out comments &
114 ($v_0$) $\to$ 100 ($v_{10}$) across 240 runs/arm; paired HL $\Delta = 0.00$ ($n=120$ issues, mean over two trajectories; 68 unchanged); 95\% CI \texttt{sqlglot} [0.00, 0.00], \texttt{dvc} [0.00, 0.25]; same-prompt \texttt{sqlglot} $v_0$ replicates differed by 15 (24 vs.\ 39). &
Inside replicate noise; no measurable effect. \\
Comment depth &
311 local correctness, 45 design, 34 requirements/spec, 4 surface of 394 ($\kappa=0.45$); held-out design comments 7/63 (11\%) $\to$ 2/44 (5\%) in \texttt{sqlglot}, 7/51 (14\%) $\to$ 5/56 (9\%) in \texttt{dvc}; paired HL $\Delta = 0.00$ ($n=38$ issues with comments in both arms). &
Sparse comments leave depth shifts inconclusive. \\
External quality &
Hidden-test resolved 62 ($v_0$) $\to$ 56 ($v_{10}$) of 240 (McNemar: 51 both, 11 $v_0$-only, 5 $v_{10}$-only, 173 neither, exact $p=0.21$, discordant-pair odds ratio $11/5 = 2.20$; 103/120 issues invariant across all 4 runs, 17 ever flip). PRs with new visible-test failures after \agent{Fix}: \texttt{sqlglot} 16 $\to$ 15 of 120, \texttt{dvc} 5/120 $\to$ 5/119 (HL $\Delta = 0.00$ [0.00, 0.00]). &
Bimodal benchmark; no detectable change (discordant pairs favor $v_0$). \\
Process signals &
Held-out \agent{Planning} \texttt{spiral} stops fell 71 $\to$ 52 of 240 ($v_0 \to v_{10}$; exploratory; $p=0.03$ unadjusted, $0.27$ Holm-adjusted); median tokens (1.53M $\to$ 1.57M) and wall-clock (21.5 $\to$ 22.5 min) flat. &
Inconclusive shift, unseen by the \agent{Retrospective} agent. \\
\bottomrule
\end{tabular}
\end{table*}

Three post-loop checks complemented hidden tests. First, a descriptive uptake check planted one probe rule at the start or end of the $v_{10}$ \agent{Implementation} rules of all four trajectories on the 60 held-out issues of each repository (alternating start/end) vs.\ matched no-probe $v_{10}$ runs. The file-level probe asked for \texttt{\# Changed: <description>} above the first changed line of each modified existing \texttt{.py} file; the function-level probe asked for a docstring starting with \texttt{Changed:} on every added or modified function. Because the function-level probe had only been piloted in dry runs, we added it and the replicate-B file-level runs (360 PRs) after the freeze, so that both shapes were measured on identical issues and prompts (240 PRs each). Second, a record-only visible-test check ran the full visible suite offline after \agent{Implementation} and \agent{Fix} on all 880 PRs. Third, an external LLM classifier (Jev~1.14\footnote{Decision model from \href{https://typesafe.ai/}{TypeSafe AI}}, given our frozen depth codebook) assigned each comment a depth class ($\kappa=0.45$ against 100 hand-coded comments).

\Figure{fig:chain} and \Table{tab:results} trace the signal through the loop. Governance and uptake worked as designed, but the review signal entering the loop was sparse, and no held-out measure changed measurably between $v_0$ and $v_{10}$.

\section{Lessons Learned}
\label{sec:pitfalls}

Each step of the chain where the signal thinned in \Section{sec:first-run} exposes lessons learned (LL) affecting the loop's sensor, \agent{Notary} checks, or study design.

\textit{LL 1: Signal starvation.} \agent{Retrospective} had on average only $0.45$ comments/PR to draw suggestions from. Pre-run power calculations assumed $\sim 6$ comments/PR from our earlier four-reviewer study (\texttt{Qwen3.6-35B-A3B} on 461 PRs)~\cite{heander2026codehealthpredictstokencost,heander2026codehealthreplication}. A comparison with that study suggests three contributing differences. First, merging four reviewers into one and forcing design comments to be grounded in consequences reduced design comments from $3.48$ to $0.08$/PR. Second, SWE-rebench PRs are smaller (median $26$ vs.\ $93$ lines) and comment rates rose from $0.19$ at 1--10 lines to $1.28$ above 100. Third, a stricter four-condition bar with nine exclusions curbed logic and security/performance comments ($0.35$ and $0.005$/PR). \agent{Review} commented on $42\%$ ($284/681$) of PRs that later failed hidden tests, a rate that parse errors ($1/880$), empty patches ($28/880$), or review \texttt{spiral} stops ($130/880$; these truncated reviews averaged $0.64$ comments/PR vs.\ $0.41$ for the rest) do not explain. Future studies should evaluate adding a separate design reviewer and using a dataset with more PRs to run more ($k>10$) or larger ($N>10$) learning windows.

\textit{LL 2: Blindness to process signals.} Across 880 runs, none of the 32 PRs whose \agent{Implementation} session ended in a \texttt{spiral} (16) or \texttt{explore\_stall} (16) stop resolved its issue, vs.\ $16/103$ after \texttt{write\_drought} and $181/738$ after normal completion. Held-out \agent{Planning} \texttt{spiral} stops fell from 71 to 52 of 240 although \agent{Retrospective} never saw stop reasons or cost. The drop is exploratory and does not survive correction (unadjusted $p=0.03$, Holm-adjusted $p=0.27$ over ten stage and stop-reason cells), but it suggests that learned rules could improve process signals that the loop does not observe. Giving \agent{Retrospective} and \agent{Notary} guard stops, tool-error rates, and stage token costs directly addresses part of signal starvation while preserving blindness to held-out measures.

\textit{LL 3: Outcome-blind rule admission.} Checking provenance, contradiction, evidence, and specificity statically left \agent{Notary} blind to effects on outcomes, which mirrors the poor text-only lesson selection by agents~\cite{pham2026playbooks}. Harness$^2$ instead executes candidate and reference harnesses on the same task and has an LLM read the contrast between their trajectories and deliverables~\cite{xu2026harness2}. Adapting this, the \agent{Notary} can propose $M \ge 1$ prompt-edit variants after each window, re-run a sample of $S$ window PRs under each variant and the unchanged $v_k$, and admit a variant only when a profile of measured contrasts improves without regressions. Contrasts could include guard stops and spirals, stage token cost, tool errors, visible-test failures, test coverage, and static code-health metrics~\cite{heander2026codehealthpredictstokencost}. Review comments join this profile only with a reviewer that comments enough, such as the proposed design reviewer of LL~1. Under Run~1's \agent{Review} agent, identical $v_0$ prompts on the same 119 held-out issues differed by $SD=1.20$ comments/PR, so detecting a shift of $0.2$ comments/PR (about 40\% of the mean) at one-sided $\alpha=0.05$ and 80\% power would need ${\sim}225$ paired re-runs per variant, against 10 PRs per window. Re-runs should also target the PRs whose comments a proposal cites, where the expected effect is largest.

\textit{LL 4: Nothing forced or informed rule retirement.} A 3,000-token per-stage ceiling never bound (peak 820 tokens; 46 additions, 5 merges, 4 replacements, and no deletions). Setting ceilings near seed size (e.g., $\sim400$ tokens/stage) would force merging and retirement so context stays bounded and auditable. Retirement also lacked evidence: the ledger recorded why each rule was admitted but not if it later reduced the issues it cited, so neither \agent{Retrospective} nor \agent{Notary} could tell if a rule worked. Recording each rule's predicted and observed effect, as optimizer memory does for harness edits~\cite{yu2026sim}, would give \texttt{delete} proposals evidence.

\textit{LL 5: Short runs learn domain practice, not repository knowledge.} Most \texttt{dvc} rules were generic (19 of 22), and most \texttt{sqlglot} rules (22 of 26) were project-specific mainly in vocabulary, restating general practice such as scoping changes to one dialect. Seeding $v_0$ with a contributor guide or textbook baseline~\cite{tomar2026textbook} and running 30--50 windows could show whether longer runs learn repository knowledge.

\textit{LL 6: Admission does not ensure uptake.} Under identical issues and prompts, agents followed a once-per-file comment rule in $65\%$ of files ($42\%$ of PRs fully) but a per-function docstring rule in only $16\%$ of functions ($2\%$ of PRs), although both named one concrete action. The uptake of the admitted rules themselves is unknown: $65\%$ of them say ``verify,'' which leaves no trace in the diff. The \agent{Notary} should prefer rules whose compliance is observable in the PR, so that their uptake can be measured.

\subsection{Threats to Validity}
Using one local LLM and two Python repositories limits generalization to frontier LLMs and other codebases. The frozen \agent{Review} agent served as both sensor and outcome measure, although hidden tests and the \agent{Notary}'s prompt-leakage check reduce this coupling. One LLM both proposed and judged rules (\Section{sec:loop}). The held-out set limits detectable effects: 103 of 120 issues gave the same test outcome in all four runs, and same-prompt $v_0$ replicates differed by 15 comments. The two uptake probes differ in the number of actions they require per PR, and each issue ran once per arm. The function-level probe, the replicate-B probe runs, and the guard-stop contrast were added after the freeze and are exploratory. Depth classes come from an external classifier with moderate agreement ($\kappa=0.45$), and rule codes rest on consensus between two authors rather than a measured agreement statistic.

\section{Conclusion and Vision}
\label{sec:directions}

Run~1 showed that a \agent{Notary} can keep a self-editing loop auditable, rejecting 34 of 89 proposals on provenance and evidence, yet admission on cited evidence left no measurable trace in held-out outcomes. We envision governed self-learning agentic loops in which every learned rule is a testable hypothesis. A proposal would name the signal its rule should change, such as a recurring review comment, a guard stop, or a failing visible test. The \agent{Notary} would admit the rule only when targeted re-runs of the cited PRs show that signal receding, and retire it when the effect fades. The ledger would then record not only why each rule entered but whether it still works, giving maintainers an auditable record to review, veto, or adopt into their own contributor guides. Evaluating this vision calls for a larger study with identical-prompt, contributor-guide, and placebo-loop controls~\cite{tomar2026textbook} on tasks with mixed baseline outcomes.

\section*{Acknowledgments}
This work was supported by the Swedish strategic research environment ELLIIT, the Wallenberg AI, Autonomous Systems and Software Program (WASP) funded by the Knut and Alice Wallenberg Foundation, Sweden's Innovation Agency (grant no. 2026-00097), and the Competence Centre {\sc NextG2Com} funded by the VINNOVA program for Advanced Digitalisation with grant number 2023-00541. The authors used Claude Code (Anthropic) with Opus 5.5, Sonnet 5.5, and Haiku 5.5 to implement and review the study's harness, analysis, and figure code under the authors' specification and review, and to draft and edit manuscript text. The authors reviewed and revised all manuscript text, verified all code with tests, checked every reported number against the data, and take full responsibility for the content.

\section*{Data Availability}
Pre-registration package at \href{https://doi.org/10.5281/zenodo.23978257}{DOI:10.5281/zenodo.23978257}. Source code and data at \href{https://doi.org/10.5281/zenodo.24338056}{DOI:10.5281/zenodo.24338056}.

\bibliographystyle{IEEEtran}
\bibliography{references}

\end{document}